\documentclass[twocolumn]{raa_twocolumn}           

\usepackage{graphicx,times}
\usepackage{natbib}
\usepackage{amssymb,amsmath}
\bibpunct{(}{)}{;}{a}{}{,}

\usepackage[pagebackref=true]{hyperref}

\begin{document}

\title{Testing the TDE Outflow Model for the Bipolar Sgr A Lobes at the Galactic Center}

   \setcounter{page}{1}

\author{Sida Li\inst{1,2}, Fulai Guo\inst{1,2,3}}

\institute{Key Laboratory for Research in Galaxies and Cosmology, Shanghai Astronomical Observatory, Chinese Academy of Sciences, 80 Nandan Road, Shanghai 200030, China; \textit{fulai@shao.ac.cn}\\
    \and
    School of Astronomy and Space Science, University of Chinese Academy of Sciences, 19A Yuquan Road, Beijing 100049, China\\
    \and
    Tianfu Cosmic Ray Research Center, 610000 Chengdu, Sichuan, China
}


\abstract{
Sgr A lobes are a pair of 15-pc-sized bipolar bubbles with co-aligned major axes perpendicular to the Galactic plane found in X-ray and radio observations of the Galactic center (GC). Their elusive origin is a vital ingredient in understanding the ongoing high energy processes at the GC. Here we perform a suite of hydrodynamic simulations to explore the tidal disruption event (TDE) outflow model for the origin of the Sgr A lobes. By following the outflow evolution in the circumnuclear medium, we find that TDE outflows naturally produce bipolar lobes delimited by forward shocks, and the dense postshock shell contributes significantly to the lobe's X-ray emission. Our fiducial run reproduces the morphology, density, temperature, and X-ray surface brightness distribution of the observed Sgr A lobes reasonably well. The current lobe age is $\sim 3300$ yr. Our model further predicts that the uplifted wake flow breaks through the ejecta-induced shock front, producing a shock-enclosed head region which is relatively dim in X-rays compared to other lobe regions. Both the dim head region and the predicted limb-brightening feature of the lobes are slightly inconsistent with current X-ray observations, and may be used to further test our model with more sensitive future X-ray observations. While light narrow jets and massive wide winds from TDE events usually do not reproduce the observed oval-shaped morphology of the lobes, the TDE outflow in our fiducial run is massive and yet narrow. Whether it is a jet or wind remains uncertain and future simulations covering both the outflow acceleration region and its pc-scale evolution would be very helpful in determining whether the Sgr A lobes indeed originate from a pair of TDE jets or winds.
}

   \authorrunning{Li \& Guo}            
   \titlerunning{Testing the TDE Outflow Model for the Bipolar Sgr A Lobes}  
   \maketitle

\section{Introduction} \label{sec:intro}

Observations indicate that there exists a supermassive black hole (SMBH), denoted as Sagittarius A* (Sgr A*), located at the center of the Galactic Center (GC), with a mass of about $4\times 10^6 $ solar mass \citep{Genzel2010,Ghez2012}. As the closest SMBH, Sgr A* provides an excellent site for some extreme physical processes, some of which may inject a large amount of mass and energy into its environment. Although it is quite dormant currently, its past activities may thus have significantly shaped the gas distributions in the Milky Way on various temporal and spatial scales, potentially resulting in the formation of the well-known 10-kiloparsec-sized \textit{Fermi} bubbles in the inner Galaxy as an example \citep{su2010,Guo2012a,Mou2014,ruiyu2020,Yang2022}. 

On the smaller parsec scales around Sgr A*, recent X-ray and radio observations reveal a pair of 15-parsec-sized bipolar lobes of oval shapes with co-aligned major axes oriented perpendicular to the Galactic plane \citep{Morris2003,Markoff2010,Heard2013,Ponti2015,Ponti2019,Zhao2016}. These two lobes, often denoted as the Sgr A lobes, connect at and appear symmetrically around Sgr A*, suggesting that they may originate from a small region around it. Both Chandra and XMM-Newton have measured X-ray emissions \citep{Morris2003,Heard2013,Ponti2015,Ponti2019} from these lobes, indicating that thermal emissions from the diffuse hot gas with temperatures of about $0.7-1$ keV contribute significantly or even dominantly. X-ray surface brightness and the derived gas pressure have strong gradients along the Galactic latitude, indicating that the gas in the lobes are outflowing away from the Galactic plane. X-ray images of the lobes show sharp and well-defined edges, suggesting that Sgr A lobes may be enclosed by forward shocks and a thin postshock shell may contribute significantly to the X-ray emissions of the lobes.  

As discussed in \citet{Markoff2010} and \citet{Clavel2019}, there are two potential classes of the formation mechanism for the bipolar Sgr A lobes: quasi-continuous outflows, including the collective stellar wind from the central cluster of young massive stars and steady outflows from recent accretion activities of Sgr A*, and an explosive outburst such as a supernova near Sgr A* or a recent tidal disruption event (TDE) by Sgr A*. The collective stellar wind from the central stellar cluster, whose age is about 6 Myrs \citep{paumard06}, is expected to reach much larger distances than the current size of the lobes. Steady outflows from Sgr A*, if continuously fed by stellar winds from the central stellar cluster, is also expected to have reached much larger distances during the past 6 Myrs. \citet{Yalinewich2017} and \citet{Ehlerova2022} performed hydrodynamic simulations of supernova evolution near Sgr A* and found that a supernova provides enough energy to form Sgr A lobes and the supernova remnant (SNR) can reach $15$ parsecs within several thousands years. However, a supernova explosion typically produces only one single lobe, which may be deformed in shape by stellar winds or molecular clouds, but could not explain the well-defined morphology of the two bipolar lobes around Sgr A*.  

In this work, we focus on the TDE outflow model for Sgr A lobes. A SMBH can capture a star that moves too close to it and tear apart the star. Roughly half of the debris of this disrupted star then falls back to the central SMBH \citep{Hills1975,Rees1988} and forms a super-Eddington accretion flow, which generates strong winds or jets as shown in many simulations \citep{Dai2018,Curd2019,Bu2022}. A TDE outflow carries comparable or even larger energy than a supernova explosion, and is thus also regarded as a possible origin for the Sgr A lobes. Sub-parsec-scale hydrodynamic simulations of TDE jets in the circumnuclear medium (CNM) by \citet{Colle2012} and \citet{Mimica2015} confirm that a pair of TDE jets indeed produce two bipolar lobes enclosed by forward shocks. A series of successive TDE outflows have also been invoked by \citet{Cheng2011} and \citet{Ko2020} to explain the origin of the Fermi bubbles.

Here we perform the first hydrodynamic simulations of the TDE outflow model for the Sgr A lobes. Compared to \citet{Colle2012} and \citet{Mimica2015}, our simulations cover a much larger spatial region, from sub-parcsec to tens of parsec scales, and are specially tailored for the GC environment in the Milky Way. The long-duration large-scale evolution of a TDE outflow is itself an interesting topic. Compared with continuous active galactic nucleus (AGN) jets, a TDE outflow injects less energy within a much shorter timescale ($\sim$ several months), and may thus not be able to sustain a fast advancing phase over an extended period of thousands of years or longer. We performed a large set of simulations, where the parameters of the TDE outflows (winds or jets) are chosen to fit X-ray observations of the Sgr A lobes. To test the TDE outflow model, we compare the morphology and X-ray surface brightness distribution of the resulting lobes with X-ray observations. In Section \ref{sec2}, we describe the basics of our model and numerical setup. Simulation results and comparisons with observations are presented in Section \ref{sec3}. In Section \ref{sec4}, we summarize our main conclusions and discuss some interesting features and uncertainties in our model.

\section{Methodology} \label{sec2}

\subsection{Numerical Setup}

Assuming axisymmetry around the Galactic rotation axis, we solve the basic hydrodynamic equations (e.g., \citealt{guo18}) in $(r, \theta)$ spherical coordinates using the widely-used ZEUS-MP code (\citealt{hayes06}). As the two Sgr A lobes are co-aligned perpendicular to the Galactic plane and connect at Sgr A*, we assume that the TDE outflow is ejected axisymmetrically along the Galactic rotational axis from the GC. As the lobe age is much shorter than the gas cooling time, we neglect radiative cooling in our simulations. The gas is assuming to be ideal following $P=k_{\rm B}T\rho/(\mu m_{\mu})=k_{\rm B}Tn$, where $k_{\rm B}$ is Boltzmann's constant, $m_{\mu}$ is the atomic mass unit, $\mu=0.61$ is the mean molecular weight per particle, and $P$, $T$, $\rho$, $n$ are the gas pressure, temperature, density, and number density, respectively.

To solve the outflow evolution in the CNM of the Milky Way, we generate 600 logarithmically-spaced grids in the $r$ direction and 400 uniform grids in the $\theta$ direction. The radial grid covers a range from $r_{\rm min}=0.1$ pc to $r_{\rm max}=30$ pc from Sgr A*, with $\Delta r_{i+1}/\Delta r_{i}=1.007$. Each radial grid is $0.7\%$ larger than the preceding one, and the innermost grid has the best spatial resolution of $1.002\times10^{16}$ cm. We focus on the evolution of one single TDE outflow in one hemisphere, and thus the $\theta$ grid covers a range from $\theta=0^{\circ}$ to $90^{\circ}$. Along the $\theta$ direction, we adopt reflective boundary conditions at both the inner and outer boundaries. Along the $r$ direction, we adopt outflow boundary conditions at both the inner and outer boundaries except that during the active outflow phase we use inflow boundary conditions at the inner boundary to initialize the TDE outflow (see Sec. \ref{sec:outflow} for details).

\subsection{Gravitational Potential and Initial Conditions}

As the current size (major axis along the Galactic rotation axis) of the Sgr A lobes is about 15 pc, we investigate the TDE outflow evolution until the triggered forward shock reaches a largest distance of about 15 pc. Within the central 20 pc from Sgr A*, the mass is dominated by Sgr A* and the nuclear star cluster (NSC). Thus in our simulations, we consider a static gravitational potential contributed by these two components. For Sgr A*, we adopt the Newtonian gravitational potential with $M_{\rm bh} = 4\times 10^6 ~M_{\odot}$. For the NSC's gravitational potential, we adopt the logarithmic form proposed by \citet{stolte08}, $\Phi = 0.5v_{0}^2 \rm{ln}(R_{\rm c}^2+r^2)$, where $v_{0}=98.6$ km s$^{-1}$, and the core size $R_{\rm c}=2$ pc and $r$ could be in units of any length scale.

\begin{figure}
  \centering
  \includegraphics[width=8cm]{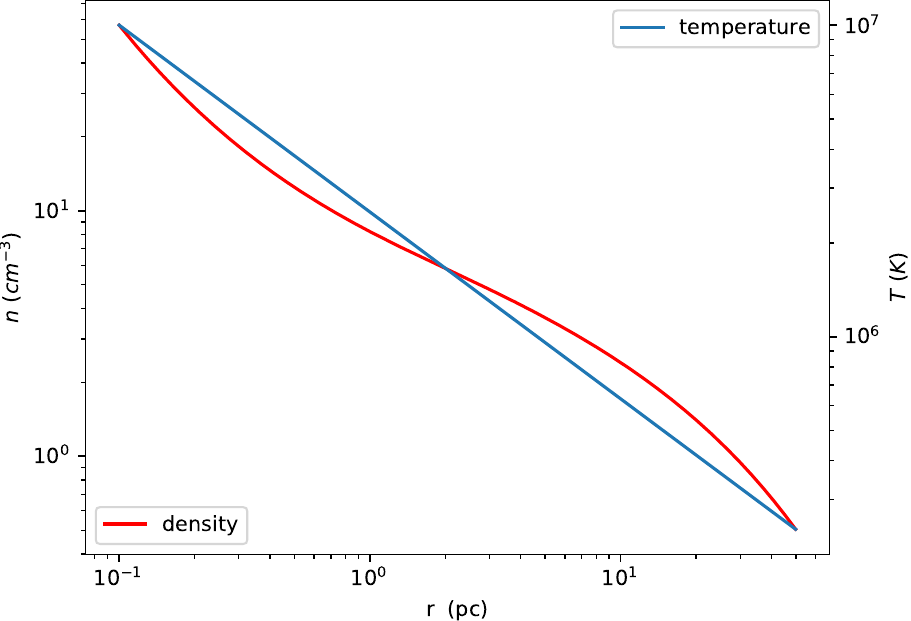}\\
  \caption{Initial gas density and temperature distributions of the CNM as a function of the distance to Sgr A*. }\label{fig1}
\end{figure}

In the GC, TDE outflows evolve in the CNM, which is shaped by the interplay between the pre-existing diffuse gas and stellar winds from the central young stellar cluster. For simplicity, we assume that the CNM gas is initially in hydrostatic equilibrium and adopt the following power-law temperature distribution,
\begin{equation}\label{eq1}
  T(r) = T_0 (\frac{r}{0.1~\rm{pc}})^{-0.6}~{,}
\end{equation}
where the inner gas temperature at $r=0.1$ pc is chosen to be $T_0 = 10\time10^7$ K. Such a gas temperature distribution is roughly consistent with the simulation predictions of collective stellar winds from the central young stellar cluster at the GC \citep{Quataert2004,Calderon2020}. The initial gas density distribution is then solved from hydrostatic equilibrium with a normalization of $n_{\rm out} = 0.5$ cm$^{-3}$ at $r = 50$ pc. The initial temperature profile and the resulting density profile are shown in Figure \ref{fig1}. The derived gas density at the inner boundary $r_{\rm min}=0.1$ pc is $55$ cm$^{-3}$, which is roughly consistent with Chandra observations of the GC \citep{baganoff03}. 

\subsection{Outflow Setup}
\label{sec:outflow}

We adopt inflow boundary conditions at $r=r_{\rm min}$ to initialize TDE outflows in our ZEUS-MP simulations. Specifically, during the active outflow phase $0\leq t \leq t_{\rm inj}$, we set inflow boundary conditions mimicking a TDE outflow at the radial inner boundaries $r=r_{\rm min}$, $0\leq \theta \leq \theta_{\rm half}$, where $\theta_{\rm half}$ is the half-opening angle of the outflow. Note that we have implicitly assumed that the outflow axis co-aligns with the Galactic rotation axis. These inflow boundary conditions will be replaced by regular outflow boundary conditions when $t > t_{\rm inj}$, where the outflow duration is set to be $t_{\rm inj}=0.5$ yr in all our runs. TDE accretion simulations often show that the velocity of a TDE outflow (jet or wind) has an angular structure \citep{Dai2018,Curd2019,Bu2022}. Here we assume that the TDE outflow velocity is mainly radial and adopt the following angular dependence of the radial outflow velocity at the inner boundary, which provides a reasonably good fit to the simulated TDE outflow distribution in \citet{Dai2018},
\begin{eqnarray}
  v_{r}(\theta) = 
     \begin{cases}
     v_0 e^{-2.5\theta/\theta_0} ~~~ &\rm{if} ~ 0\leq \theta \leq \theta_{\rm half,} \\
      0 ~~~ &\rm{if} ~ \theta > \theta_{\rm half,}
     \end{cases}
     \label{eq2}
\end{eqnarray}
where $\theta_0=1$ radian. For simplicity, we assume that the outflow density $\rho_{\rm inj}$ and temperature $T_{\rm inj}$ do not vary with $\theta$.

An important characteristic of TDE outflows is their exponential decay with time. Since the fallback rate of stellar debris is proportional to $t^{-\frac{5}{3}}$, theoretical works often assume that the accretion rate has the same behavior, and so does the outflow power \citep{Mimica2015,Winter2021}. In this work, we add this temporal decay to the outflow density $\rho_{\rm inj}$,
\begin{eqnarray}\label{eq3}
  \rho_{\rm inj}(t) = 
   \begin{cases}
   \rho_0  ~~~ &\rm{if} ~ t < t_0, \\
   \rho_0 (\frac{t}{t_0})^{-5/3}  ~~~ &\rm{if} ~ t \geq t_0,
   \end{cases}
\end{eqnarray}
where $t_0=0.01$ yr. The exact value of $t_0$, which is very small compared to the TDE duration, does not affect our main results. The outflow temperature $T_{\rm inj}$ is assumed to be constant with time. In all our simulations, the outflow power is dominated by its kinetic power.

We use the term ``TDE outflows" throughout the paper and do not explicitly distinguish between jets and winds. We performed a large suite of simulations, exploring the space of outflow parameters with a focus on $\rho_{0}$ and $\theta_{\rm half}$. Large values of $\rho_{0}$ (or total injection mass $M_{\rm inj}$) and $\theta_{\rm half}$ could be considered as characteristics of TDE winds, but we note that wide-opening winds may also undergo substantial collimation during their evolution from the launching region to $r_{\rm min}=0.1$ pc.

\section{results} \label{sec3}

We have performed a large suite of simulations with different combinations of outflow parameters, and here in the following two subsections, we present the results of the fiducial run, denoted as run FR, which fit X-ray observations of the Sgr A lobes reasonably well. In run FR, the adopted outflow parameters are $\rho_0 = 9.11\times10^{-19}$ g cm$^{-3}$, $T_{\rm inj} = 1\times 10^8$ K, $v_0=7.5 \times 10^9$ cm/s, and $\theta_{\rm half}=8^{\circ}$. The underlying properties of this outflow will be further discussed in Section \ref{attempts}, where we also present the results of two additional runs, TJ for a thin jet run and WW for a wide wind run, to briefly show how our results depend on outflow parameters.

\subsection{Dynamical Evolution of TDE Outflows}

\begin{figure}
  \centering
  \includegraphics[width=8cm]{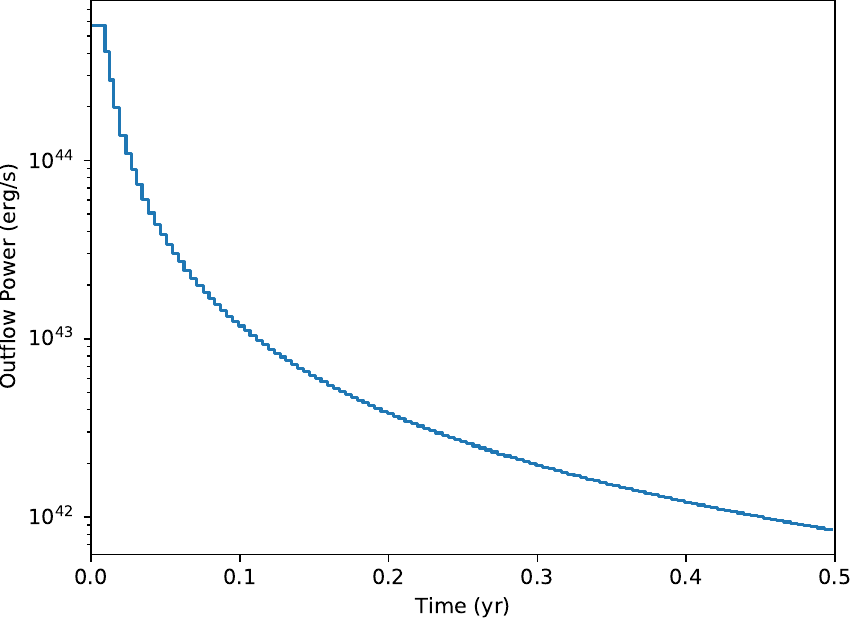}\\
  \caption{Temporal decay of the TDE outflow power (see Eq.\ref{eq3}) in our fiducial run FR, which reproduces X-ray observations of the Sgr A lobes reasonably well.}\label{fig2}
\end{figure}

\begin{figure*}
  \centering
  \includegraphics[width=\textwidth]{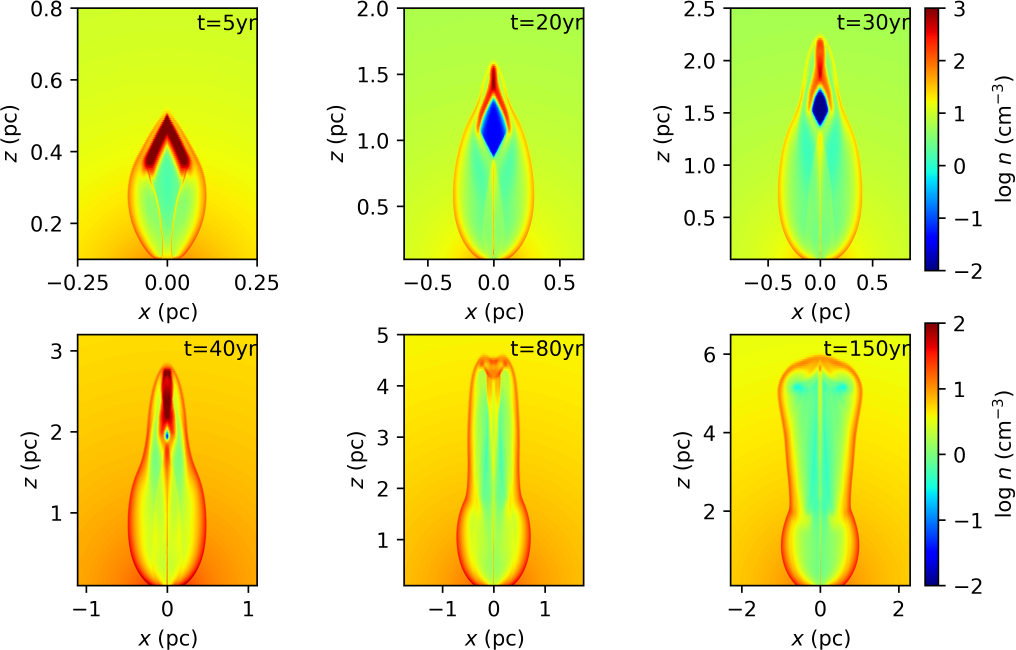}\\
  \caption{Gas density distributions at $t=5$, $20$, $30$, $40$, $80$ and $150$ yr in our fiducial run FR, showing the early time evolution of the TDE outflow in the CNM at the GC. The upper and lower colorbars on the right apply to the panels in the upper and lower rows, respectively.}\label{fig3}
\end{figure*}

The early time evolution of the TDE outflow in run FR is shown in Fig. \ref{fig3}. The outflow density is much higher than the ambient gas density in the CNM, and the outflow duration $t_{\rm inj}=0.5$ yr is also very shorter compared to the typical evolution time. The first and second panels of Fig. \ref{fig3} show that the outflow ejecta forms an arrowhead-like morphology, which lasts for a duration much longer than the outflow injection duration. One can also clearly see that the outflow induces a bow shock propagating into the CNM, forming an oval-shaped shock front. The outflow ejecta is then subject to thermal and ram pressures of the postshock CNM gas, which squeezes the ejecta gradually towards the vertical $z$ axis, as clearly seen in Fig. \ref{fig3}. At $t=40$ yr, most of the outflow ejecta has already been squeezed along the outflow ($z$) axis, forming a linear high-density structure moving towards the $+z$ direction.

As the high-density ejecta mainly move along the $z$ axis, they push the forward shock along the $+z$ direction much stronger than along the lateral directions. Consequently at $t=15$ yr as seen in Fig. \ref{fig3}, the outflow ejecta start to deform the oval-shaped forward shock at its tip, and here we simply say that they break through the oval-shaped forward shock, forming a second bow shock on top of the previous one. The second shock evolves to a much larger size than the first shock, as clearly seen in the last two panels of Fig. \ref{fig3}. At $t=80$ yr, most of the ejecta have been thermalized in a hotspot at the outflow head, presumably by a reverse shock just below the hotspot.

\begin{figure}
  \centering
  \includegraphics[width=8cm]{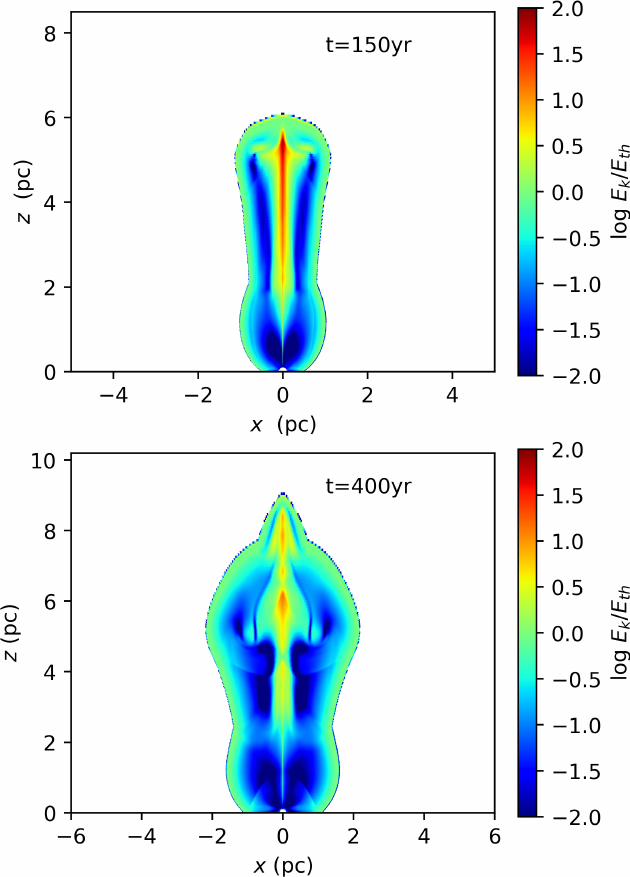}\\
  \caption{Ratios of the kinetic to thermal energy densities at $t=150$ and $400$ yr in the fiducial run FR. The linear wake flow along the $z$ axis in the top panel moves upward and breaks through the shock front, forming a third forward shock on top of the lower two.}\label{fig4}
\end{figure}

Fig. \ref{fig3} also clearly shows that a wake flow is uplifted by the outflow, forming a long linear filament along the $z$ axis (see \citealt{guo18}, \citealt{Duan2018}, \citealt{duan24} for extensive studies of wake flows in AGN jet feedback). Fig. \ref{fig4} shows the ratios of the kinetic to thermal energy densities at $t=150$ and $400$ yr. At $t=150$ yr, in a large fraction of the wake flow (mainly between $z\sim 2$ and $6$ kpc), the kinetic energy density is much higher than the thermal energy density. As the bottom panel of Figure \ref{fig4} shows, the wake flow moves upward along the $z$ axis and breaks through the second bow shock, forming a third forward shock on top of the second one. A three-layer shock structure thus forms at $t=400$ yr, as clearly shown in the bottom panel of Figure \ref{fig4}.
 
\begin{figure}[h!]
 \centering
   \includegraphics[width=8cm]{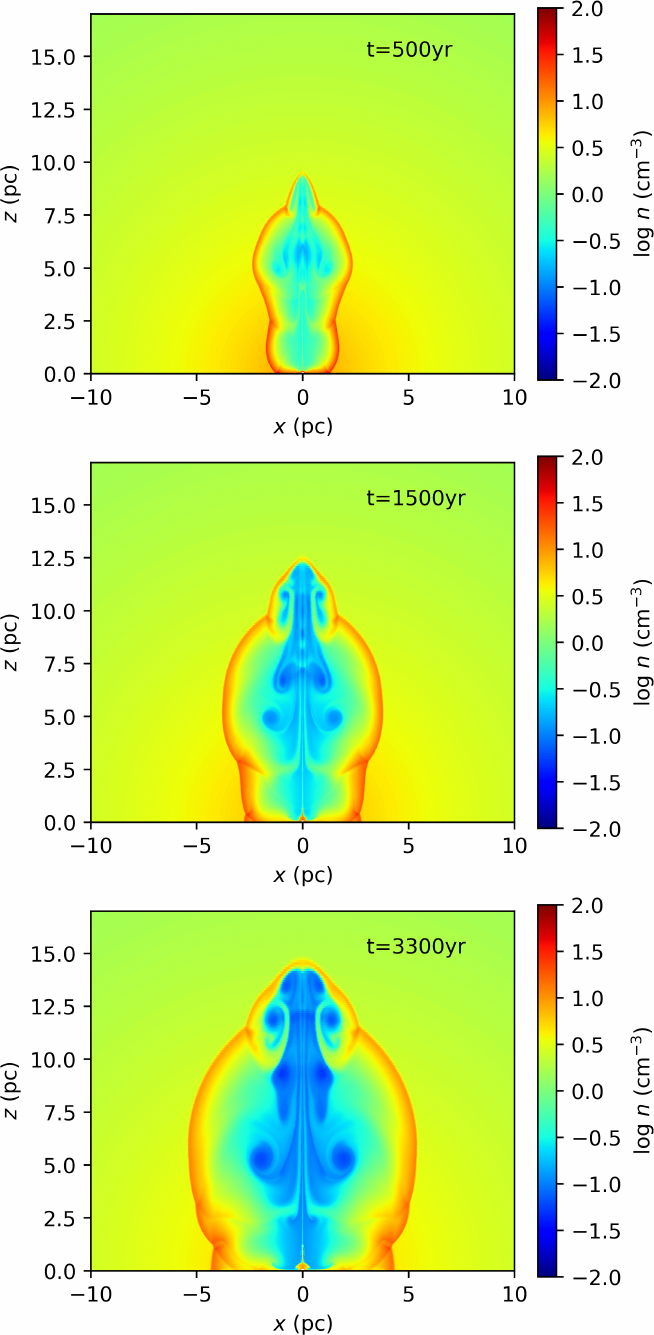}
\caption{Gas density distributions at $t=500$, $1500$, and $3300$ yr in our fiducial run FR, clearly showing the evolution of the three-layer shock structure. A large gaseous lobe naturally forms, and is delimited by the shock fronts and postshock high-density shells.}\label{fig5}
 \end{figure}

\begin{figure}[h!]
 \centering
   \includegraphics[width=8cm]{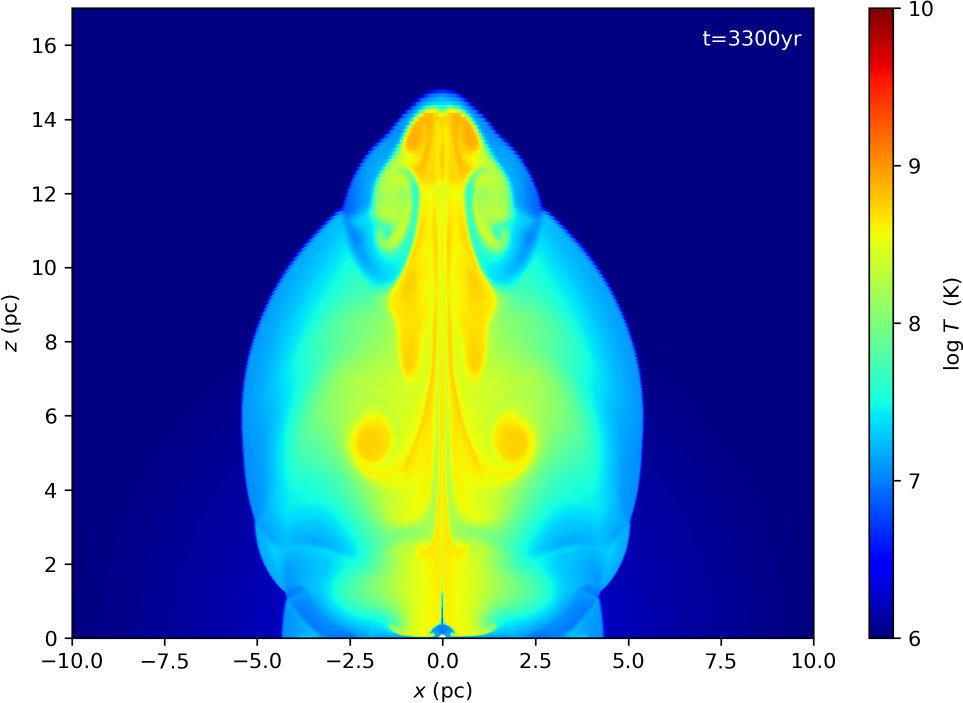}
\caption{Gas temperature distribution at $t = 3300$ yr in our fiducial run FR. }\label{fig6}
 \end{figure}

The three-layer shock structure continues to propagate to larger distances, forming a large gaseous lobe delimited by the shock fronts and the high-density postshock shells, as clearly seen in Figure \ref{fig5}. At $t=3300$ yr, the shock front reaches a height of about $15$ pc from the Galactic plane and the morphology of the lobe resembles the Sgr A lobes observed in X rays. The final gas density distribution at $t=3300$ yr is shown in the bottom panel of Figure \ref{fig5}, and the corresponding temperature distribution is shown in Figure \ref{fig6}. The gas in the simulated Sgr A lobe is concentrated within the dense shells swept up by the forward shocks, and inside the lobe, there is a large low-density high-temperature ($\sim 10^{8}$ K) cavity. Therefore, the dense shells are expected to dominate the thermal X-ray emissions from the Sgr A lobes, as further investigated in Section \ref{sec-emission}. 

We note that the final shock fronts delimiting the lobe contain three layers: the bottom forward shock inducing by the outflow injection, the middle shock triggered by the squeezed outflow ejecta breaking through the bottom shock along the outflow axis, and the top shock induced by the wake flow breaking through the middle shock along the outflow axis. 

\subsection{Comparison with X-Ray Observations}\label{sec-emission}

In this subsection, we compare the results of our fiducial run FR with X-ray observations, particularly the observed morphology and X-ray surface brightness distribution of the Sgr A lobes. To this end, we adopt the $2-4.5$ keV X-ray emissivity $\epsilon (T,Z)$ from the Astrophysical Plasma Emission code (APEC; \citealt{smith2001}) plasma model with the PyAtomDB package \footnote{https://atomdb.readthedocs.io/en/master/index.html}, where the gas metallicity $Z$ is chosen to be the solar metallicity. We assume that the hot gas is optically thin and in collisional ionization equilibrium. The synthetic $2-4.5$ keV X-ray surface brightness may be written as 
\begin{equation}\label{eq4}
  I(x,z) = \frac{1}{4\pi} \int n_{\rm e} n_{\rm H} \epsilon(T,Z) dy ~~ \text{erg s}^{-1}~\text{cm}^{-2}~\text{sr}^{-1} ~{,}
\end{equation}
where the integration is performed along the $+y$ direction from $y=-10$ pc to $10$ pc, covering the whole shock-enclosed lobe. To perform this integration, we map the gas density and temperature distributions in our spherical grid to a newly-constructed 3-dimensional Cartesian grid with a uniform grid size of $0.05$ pc.

We compare our synthetic X-ray surface brightness distribution at $t=3300$ yr in run FR with XMM-Newton X-ray observations of the Sgr A lobes in \citet{Heard2013}, where the X-ray surface brightness is given in units of ``counts/20 ks/pixel". To convert our synthetic X-ray surface brightness to this observational units, we follow \citet{Heard2013} to take $4\times 4$ arcsecond$^2$ for each pixel and adopt an approximate XMM-Newton effective area of $A_{\rm eff} = 1000$ cm$^2$, an average photon energy of $3$ keV in the $2-4.5$ keV band, and a distance of $8$ kpc to the GC. Considering obscuration by the intervening neutral materials, we further assume that the ratio of the obscured (observed) flux to the unobscured flux in this X-ray band is $15\%$, corresponding to a neutral hydrogen column density of $N_{\rm HI}\sim 7-10 \times 10^{22}$ cm$^{-2}$ (see Table 5 in \citealt{Ponti2015}). This is roughly consistent with the neutral hydrogen column density ($N_{\rm HI}\sim 6.3-8.0 \times 10^{22}$ cm$^{-2}$) derived from modelling of the X-ray emission in \citet{Heard2013}, but lower than the value derived from dust emission \citep{Ponti2015}, which has large uncertainties associated with the dust-to-HI ratio. 

\begin{figure}
  \centering
  \includegraphics[width=8cm]{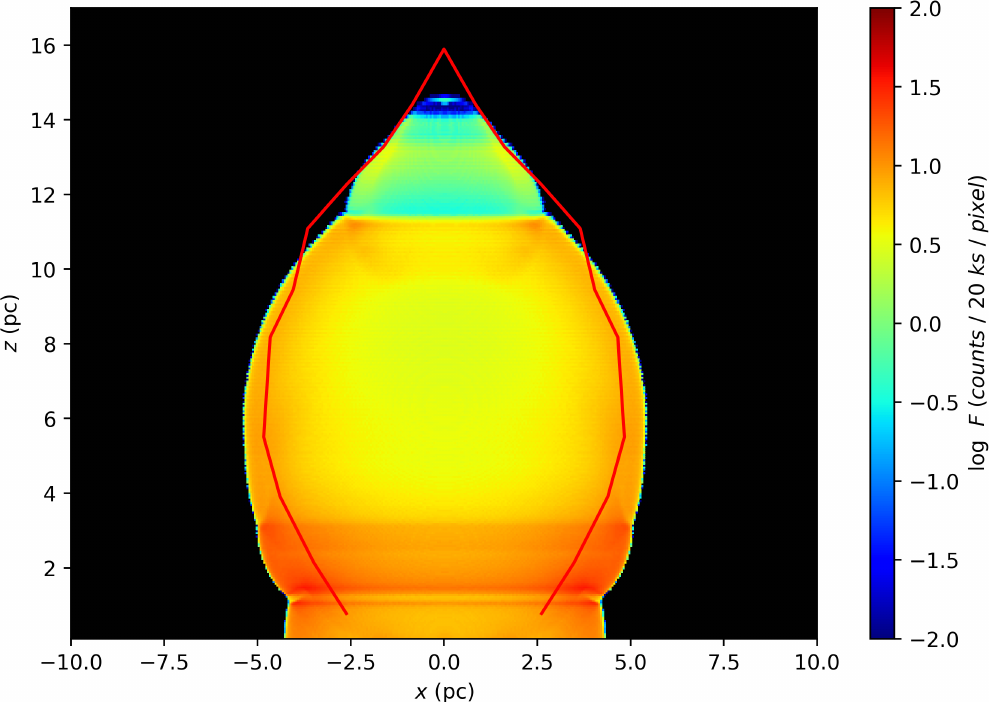}\\
  \caption{Synthetic $2-4.5$ keV X-ray surface brightness distribution at $t = 3300$ yr in our fiducial run FR. The morphology of the simulated lobe fits quite close to the observed Sgr A lobes delimited by the red line \citep{Ponti2015}. The ratio of the obscured flux to the unobscured flux is assumed to be $15\%$, giving the X-ray surface brightnesses of the simulated Sgr A lobe roughly consistent with observations \citep{Heard2013}. }\label{fig7}
\end{figure}

Figure \ref{fig7} shows the calculated synthetic X-ray surface brightness distribution at $t = 3300$ yr in our fiducial run FR. The morphology of the simulated X-ray lobe fits quite close to the observed Sgr A lobes in \citet{Ponti2015}, whose outer edges are depicted by the red line in Fig. \ref{fig7}. Since there are complicated features (possibly due to absorption) in the western part of the observed northern Sgr A lobe, here we draw the left half of the red line according to the observed eastern outer edge and then draw the right half from reflection of the left half. The simulated X-ray surface brightnesses of the lobe are of the same order of magnitude compared with X-ray observations of \citet{Heard2013}, and our simulated lobe enclosed by shock fronts naturally has sharp edges as in observations. At the lobe base at low latitudes, the simulated lobe is slightly broader than the observed Sgr A lobes, which may be due to the complex distributions of cold and hot CNM gas near the GC not accurately captured in our model. 

\begin{figure}
  \centering
  \includegraphics[width=8cm]{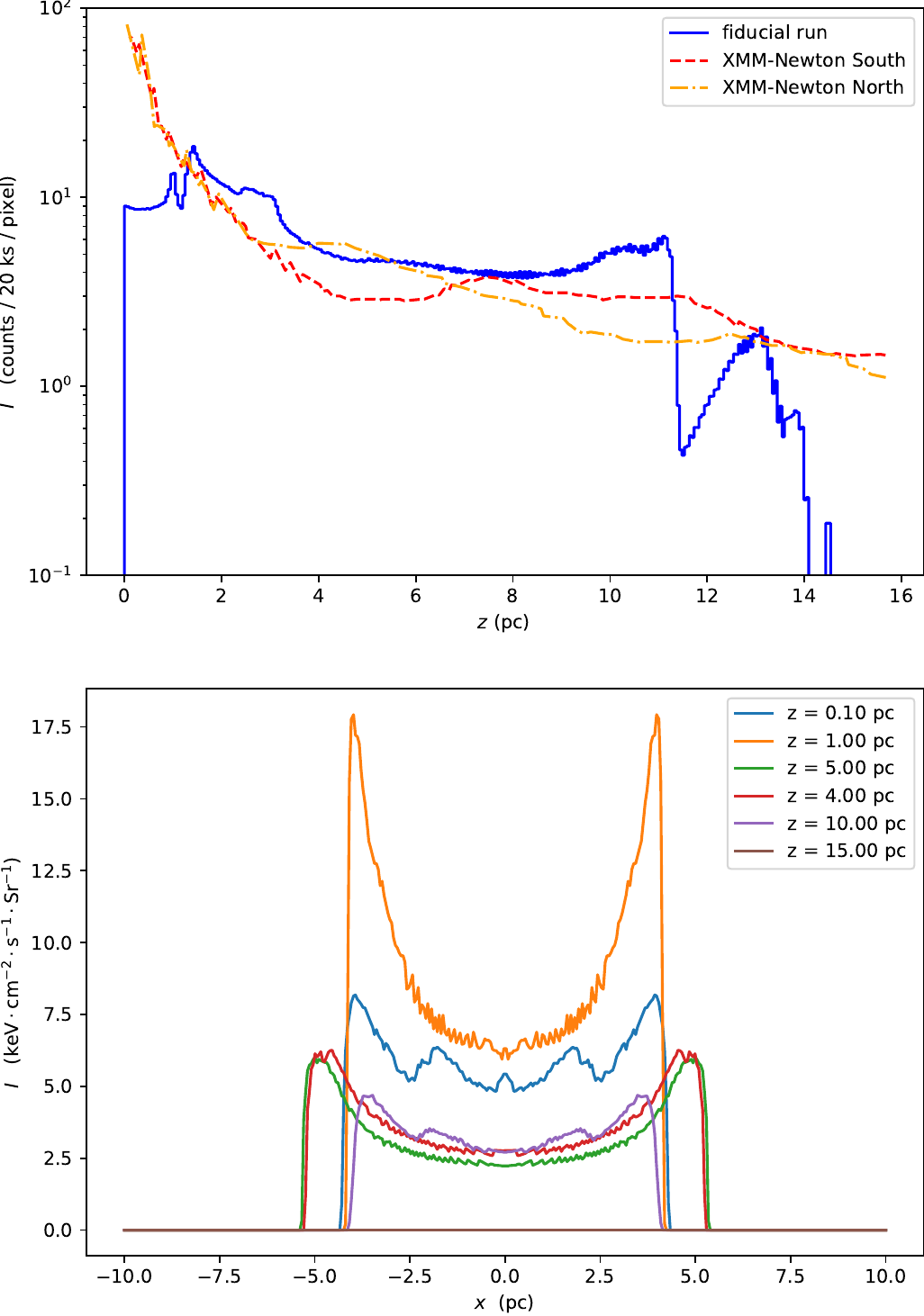}\\
  \caption{Top: Latitudinal variation of the synthetic X-ray surface brightness within a conical region with a half-opening angle of $20^{\circ}$ around the outflow axis as in \citet{Heard2013}. The dashed and dot-dashed lines correspond to XMM-Newton observations of the same conical region in the southern and northern Sgr A lobes by \citet{Heard2013}, respectively. Bottom: Longitudinal variations of the synthetic X-ray surface brightness at various heights from the Galactic plane. The limb-brightening feature is clearly seen from our simulated lobe. }\label{fig8}
\end{figure}

To compare our simulations with observations more quantitatively, we calculate the latitudinal variation of the synthetic X-ray surface brightness within a conical region with a half-opening angle of $20^{\circ}$ around the outflow axis as in \citet[Figs. 3 and 4 therein]{Heard2013}, and the result is shown in the top panel of Fig. \ref{fig8}. As seen clearly, both the observed \citep{Heard2013} and simulated X-ray surface brightnesses decrease with Galactic latitude, and are roughly consistent with each other within $\sim 11$ pc. At $z \gtrsim 11$ pc, the head of our simulated lobe is mostly dimmer than observations \citep{Heard2013}. As seen in Figure \ref{fig5}, this head region is enclosed by the top shock induced by the wake flow breaking through the middle shock along the outflow axis, and is thus an important lobe feature in our TDE outflow model. 

Our model further predicts that the lobe is limb-brightened, as seen in the bottom panel of Fig. \ref{fig8}, which shows the longitudinal variations of the synthetic X-ray surface brightness at various heights from the Galactic plane. Although such a limb-brightening feature is also mentioned in X-ray observations by \citet{Ponti2015}, the difference in X-ray surface brightness between the edges and interiors of the simulated lobe seems to be substantially larger than in observations \citep{Heard2013,Ponti2019}. Both the dim lobe head and the limb-brightening feature are key predictions of our TDE outflow model, which may be used to further test our model with more sensitive X-ray observations in the future.

With the X-ray calculations above, we further calculate the $2-4.5$ keV emission weighted average gas temperature and density within our simulated Sgr A lobe, using the following formulae:
\begin{equation}\label{eq5}
  T_{\rm ave} = \frac{\int n_e n_H \epsilon(T,Z) T dV}{\int n_e n_H \epsilon(T) dV} ~{,}
\end{equation}
\begin{equation}\label{eq6}
  n_{\rm ave} = \frac{\int n_e n_H \epsilon(T,Z)n dV}{\int n_e n_H \epsilon(T) dV}~{.}
\end{equation}
At $t = 3300$ yr in our fiducial run FR, the average gas temperature and density are $T_{\rm ave} = 1.48$ keV and $n_{\rm ave} = 8.32$ cm$^{-3}$, respectively. The average lobe temperature in our model is slightly higher than those ($T\sim 1$ keV) measured by \citet{Heard2013} and \citep{Ponti2019}, but lower than $T\sim 2$ keV in \citet{Markoff2010}. The average lobe density in our model is consistent with \citet[$n\sim 4-10$ cm$^{-3}$]{Heard2013}, but slightly higher than the values derived in \citet[$n\sim 3-4$ cm$^{-3}$]{Ponti2019}, which assume a rather uniform gas distribution within the lobes.  

In summary, our fiducial TDE outflow model (run FR) reproduces the observed morphology, density, temperature, and X-ray surface brightness distribution of the bipolar Sgr A lobes reasonably well. The age of the lobes is estimated to be $t \sim 3300$ yr in our model. Our model further predicts that the lobes are substantially limb-brightened and the lobe head is relatively dim, which seem not to be consistent with current X-ray observations and can be used to further test our model with future more sensitive X-ray observations. 

\subsection{Additional Jet and Wind Runs}\label{attempts}

In our fiducial run FR, the total injected mass and energy by one single outflow are  $M_{\rm inj} = 0.25M_{\odot}$ and $E_{\rm inj} = 4\times10^{50}$ erg, respectively. The total injected mass and energy by the two bipolar outflows should double these values. While the small half-opening angle $\theta_{\rm half}=8^{\circ}$ of this outflow may be more consistent with a jet than a wide wind, the total injected mass may be too large for a jet from a TDE of a solar-mass star. To further investigate this issue, here we present the results of two representative runs with similar amounts of energy injection, a thin jet run TJ and a wide wind run WW.

\begin{figure}
  \centering
  \includegraphics[width=8cm]{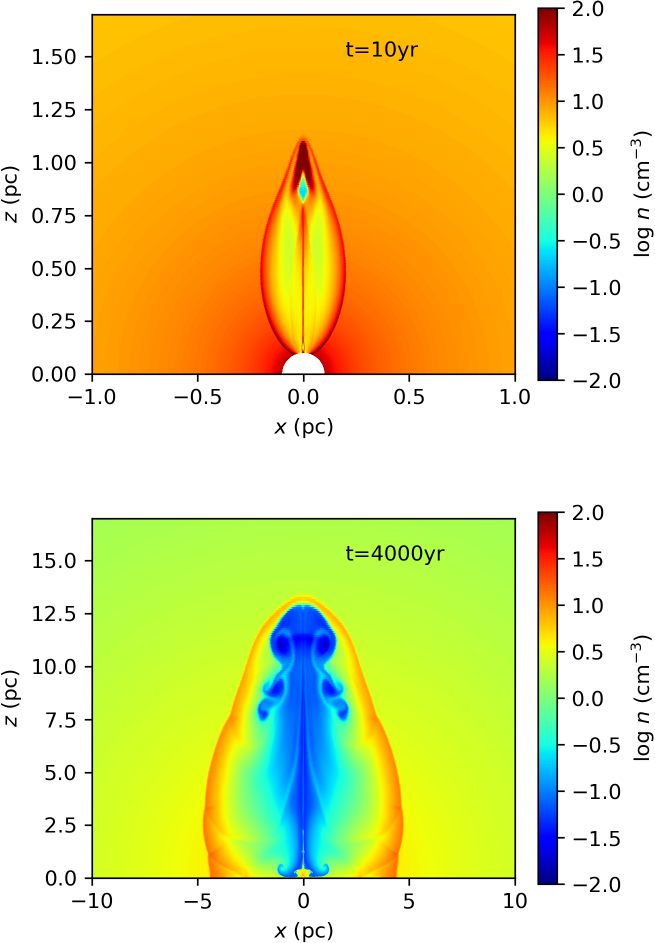}\\
  \caption{Gas density distributions at $t = 10$ and $4000$ yr in the thin-jet run TJ. The resulting lobe is bottom-wide at $t=4000$ yr.}\label{fig9}
\end{figure}

In the thin-jet run TJ, the adopted outflow parameters are $\rho_0 = 7\times10^{-19}$ g cm$^{-3}$, $T_{\rm inj} = 10^{9}$ K, $\theta_{\rm half} = 3^{\circ}$, $t_{\rm inj} = 0.5$ yr, and peak velocity $v_0 = 1.0\times10^{10}$ cm/s. The total injected mass and energy by one single outflow are $M_{\rm inj} = 0.04M_{\odot}$ and $E_{\rm inj} = 1.7\times10^{50}$ erg, respectively. Compared to run FR, the outflow in run TJ is less massive, but has higher velocities and lower half-opening angles. The outflow evolution in run TJ is quite similar to that in run FR. The thin-jet ejecta also experiences collimation towards the outflow axis by the postshock gas. A second bow shock forms by breaking the first shock at $t=10$ yr, as seen in the top panel of Fig. \ref{fig9}, which shows the gas density distribution at this time. With similar amount of energy injection as the outflow in run FR, the thin jet in run TJ has lower momentum, but receives higher ram pressure due to its higher velocity. Therefore, the thin jet decelerates faster and takes longer time for the shock tip to reach a height of $15$ pc, as seen in the bottom panel of Fig. \ref{fig9}. The resulting shock-enclosed lobe is bottom-wide, inconsistent with the oval-shaped Sgr A lobes. 

\begin{figure}
  \centering
  \includegraphics[width=8cm]{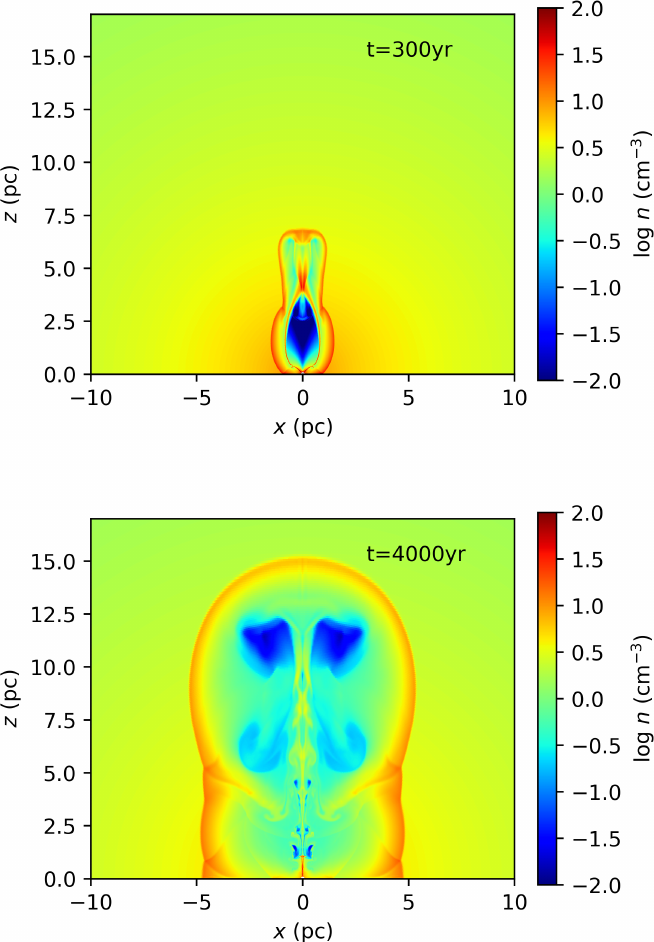}\\
  \caption{Gas density distributions at $t = 300$ and $4000$ yr in the wide-wind run WW. The ``cylindrical" lobe reaches a height of $15$ pc at $t = 4000$ yr. }\label{fig10}
\end{figure}

In the wide-wind run WW, the adopted outflow parameters are $\rho_0 = 3\times10^{-18}$ g cm$^{-3}$, $T_{\rm inj} = 10^{7}$ K, $\theta_{\rm half} = 60^{\circ}$, $t_{\rm inj} = 0.5$ yr, and peak velocity $v_0 = 3\times10^{9}$ cm/s. The total injected mass and energy by one single outflow are $M_{\rm inj} = 4.9M_{\odot}$ and $E_{\rm inj} = 2.9\times10^{50}$ erg, respectively. The wind evolution in this run is shown in Fig. \ref{fig10} and at $t=4000$ yr, the shock tip reaches a height of $15$ pc from the Galactic plane. However, the shock-enclosed lobe at this time is ``cylindrical", also inconsistent with the oval-shaped Sgr A lobes. Due to the angular decline in the outflow velocity distribution and the ram and thermal pressures from the postshock ambient gas, the outflow ejecta is collimated and stretched along the outflow axis. While a part of the ejecta is thermalized within the outflow head region (hotspot), the rest forms a long filament along the outflow axis, which suppresses the formation of linear wake flows seen in run FR.  

Both the representative thin-jet and wide-wind runs discussed above could not reproduce the oval-shaped morphology of the observed Sgr A lobes. On the other side, the massive, narrow outflow in our fiducial run FR reproduces the X-ray observations reasonably well, but its nature is somewhat uncertain. For a solar-mass star disrupted by a supermassive black hole, about $0.5M_{\odot}$ stellar debris falls back to the black hole, and most of them are expected to be ejected out through a wide wind. The opposing jets, if formed, may only carry a very small fraction of the stellar debris. The bipolar outflows in run FR carry a mass of $0.5M_{\odot}$, and if they are indeed jets, the progenitor star should be much more massive than one solar mass. The large ejecta mass could be naturally explained by a pair of wide winds, but they would require very efficient collimation on sub-pc scales to reach a half opening angle of $8^{\circ}$ at $0.1$ pc. Whether this is feasible requires further investigations, which is beyond the scope of the current paper.

\section{Summary and Discussions}\label{sec4}

In this paper, we use a series of hydrodynamic simulations to investigate the TDE outflow model for the the origin of the 15-pc Sgr A lobes at the GC. Our simulations follow the dynamical evolution of TDE outflows on parsec scales in the CNM and compare the morphology and X-ray surface brightness distribution of the resulting lobe with X-ray observations. Our simulated lobe is delimited by shock fronts and the dense postshock shell swept up by the shock contributes significantly to the lobe's X-ray emission. Our fiducial run FR reproduces the observed morphology, density, temperature, and X-ray surface brightness distribution of the bipolar Sgr A lobes reasonably well. In this ``best-fit" run, the current age of the lobes is $\sim 3300$ yr, the half-opening angle of the outflow at its base ($r=0.1$ pc) is $8^{\circ}$, and the total mass and energy injected by the two opposing outflows are $0.5M_{\odot}$ and $8\times10^{50}$ erg, respectively.  

Our fiducial run further predicts that the lobes are substantially limb-brightened due to X-ray emissions from the dense postshock shell. While the observed Sgr A lobes have sharp edges consistent with our prediction, the limb-brightening feature in observations is not as strong as in our simulations. In run FR, the lobe head is enclosed by a relatively weak shock front induced by the wake flow breaking through the ejecta-induced shock along the outflow axis, and is relatively dim in X-rays compared to other lobe regions, which seem not to be consistent with current X-ray observations. Both the predicted dim lobe head and the limb-brightening feature could be used to further test our model with more sensitive X-ray observations in the future.

The TDE outflow in our fiducial run is massive and narrow. Its underlying nature, i.e., whether it is a jet or wind, is somewhat uncertain. If it is a TDE jet, the progenitor star disrupted by Sgr A* should be much more massive than one solar mass. If it is a wind, it should be collimated very efficiently on sub-pc scales to reach a half opening angle of $8^{\circ}$ at $0.1$ pc. Additional simulations show that both the light, thin jet and massive, wide winds at our outflow base ($r=0.1$ pc) fail to reproduce the oval-shaped morphology of the Sgr A lobes. Future simulations self-consistently covering both the outflow acceleration near the black hole and its evolution on the parsec scales would be very helpful in determining whether a pair of thin TDE jets or wide TDE winds could produce the observed Sgr A lobes (or neither). 

In our simulations, the TDE outflows are all well collimated and stretched along the outflow axis during their evolution on the parsec scales. This can be attributed to the combined effect of both the angular decline in the outflow velocity distribution and the ram and thermal pressures from the postshock ambient gas. Such an anisotropic feature is often ignored in galaxy-scale simulations of TDE outflows and their impact on galaxy evolution (e.g., \citealt{Zubovas2019,Ko2020}), which usually assume spherical symmetric energy injection from TDE outflows. As shown in \citet{duan20}, the anisotropic nature of AGN jets leads to the production of backflows, substantially increasing the energy coupling efficiency between AGN jets and the ambient intracluster medium. Thus the anisotropic collimation effect should be carefully considered in future studies of TDE outflows.

\normalem
\section*{acknowledgements}
We thank Fangzheng Shi for helpful discussions on XMM-Newton observations, Daniel Wang for insightful discussions on the GC science, and Defu Bu for discussions on the TDE-related science. This work was supported by the Excellent Youth Team Project of the Chinese Academy of Sciences (No. YSBR-061) and Shanghai Pilot Program for Basic Research - Chinese Academy of Science, Shanghai Branch (JCYJ-SHFY-2021-013). The calculations presented in this work were performed using the high performance computing resources in the Core Facility for Advanced Research Computing at Shanghai Astronomical Observatory.

\bibliographystyle{raa}
\bibliography{ms}

\begin{thebibliography}{38}
\providecommand\natexlab[1]{#1}
\providecommand\JournalTitle[1]{#1}

\bibitem[{Baganoff} {et~al.}(2003)]{baganoff03}
{Baganoff}, F.~K., {Maeda}, Y., {Morris}, M., {et~al.} 2003, \apj, 591, 891

\bibitem[{Bu} {et~al.}(2022)]{Bu2022}
{Bu}, D.-F., {Qiao}, E., {Yang}, X.-H., {et~al.} 2022, \mnras, 516, 2833

\bibitem[{Calder{\'o}n} {et~al.}(2020)]{Calderon2020}
{Calder{\'o}n}, D., {Cuadra}, J., {Schartmann}, M., {Burkert}, A., \&
  {Russell}, C. M.~P. 2020, \apjl, 888, L2

\bibitem[{Cheng} {et~al.}(2011)]{Cheng2011}
{Cheng}, K.~S., {Chernyshov}, D.~O., {Dogiel}, V.~A., {Ko}, C.~M., \& {Ip},
  W.~H. 2011, \apjl, 731, L17

\bibitem[{Clavel}(2019)]{Clavel2019}
{Clavel}, M. 2019, in SF2A-2019: Proceedings of the Annual meeting of the
  French Society of Astronomy and Astrophysics, ed. P.~{Di Matteo},
  O.~{Creevey}, A.~{Crida}, G.~{Kordopatis}, J.~{Malzac}, J.~B. {Marquette},
  M.~{N'Diaye}, \& O.~{Venot}, Di

\bibitem[{Curd} \& {Narayan}(2019)]{Curd2019}
{Curd}, B., \& {Narayan}, R. 2019, \mnras, 483, 565

\bibitem[{Dai} {et~al.}(2018)]{Dai2018}
{Dai}, L., {McKinney}, J.~C., {Roth}, N., {Ramirez-Ruiz}, E., \& {Miller},
  M.~C. 2018, \apjl, 859, L20

\bibitem[{De Colle} {et~al.}(2012)]{Colle2012}
{De Colle}, F., {Guillochon}, J., {Naiman}, J., \& {Ramirez-Ruiz}, E. 2012,
  \apj, 760, 103

\bibitem[{Duan} \& {Guo}(2018)]{Duan2018}
{Duan}, X., \& {Guo}, F. 2018, \apj, 861, 106

\bibitem[{Duan} \& {Guo}(2020)]{duan20}
{Duan}, X., \& {Guo}, F. 2020, \apj, 896, 114

\bibitem[{Duan} \& {Guo}(2024)]{duan24}
{Duan}, X., \& {Guo}, F. 2024, arXiv e-prints, arXiv:2403.02807

\bibitem[{Ehlerov{\'a}} {et~al.}(2022)]{Ehlerova2022}
{Ehlerov{\'a}}, S., {Palou{\v{s}}}, J., {Morris}, M.~R., {et~al.} 2022, \aap,
  668, A124

\bibitem[{Genzel} {et~al.}(2010)]{Genzel2010}
{Genzel}, R., {Eisenhauer}, F., \& {Gillessen}, S. 2010, Reviews of Modern
  Physics, 82, 3121

\bibitem[{Guo} {et~al.}(2018)]{guo18}
{Guo}, F., {Duan}, X., \& {Yuan}, Y.-F. 2018, \mnras, 473, 1332

\bibitem[{Guo} \& {Mathews}(2012)]{Guo2012a}
{Guo}, F., \& {Mathews}, W.~G. 2012, \apj, 756, 181

\bibitem[{Hayes} {et~al.}(2006)]{hayes06}
{Hayes}, J.~C., {Norman}, M.~L., {Fiedler}, R.~A., {et~al.} 2006, \apjs, 165,
  188

\bibitem[{Heard} \& {Warwick}(2013)]{Heard2013}
{Heard}, V., \& {Warwick}, R.~S. 2013, \mnras, 434, 1339

\bibitem[{Hills}(1975)]{Hills1975}
{Hills}, J.~G. 1975, \nat, 254, 295

\bibitem[{Ko} {et~al.}(2020)]{Ko2020}
{Ko}, C.~M., {Breitschwerdt}, D., {Chernyshov}, D.~O., {et~al.} 2020, \apj,
  904, 46

\bibitem[{Markoff}(2010)]{Markoff2010}
{Markoff}, S. 2010, Proceedings of the National Academy of Science, 107, 7196

\bibitem[{Mimica} {et~al.}(2015)]{Mimica2015}
{Mimica}, P., {Giannios}, D., {Metzger}, B.~D., \& {Aloy}, M.~A. 2015, \mnras,
  450, 2824

\bibitem[{Morris} {et~al.}(2012)]{Ghez2012}
{Morris}, M.~R., {Meyer}, L., \& {Ghez}, A.~M. 2012, Research in Astronomy and
  Astrophysics, 12, 995

\bibitem[{Morris} {et~al.}(2003)]{Morris2003}
{Morris}, M., {Baganoff}, F., {Muno}, M., {et~al.} 2003, Astronomische
  Nachrichten Supplement, 324, 167

\bibitem[{Mou} {et~al.}(2014)]{Mou2014}
{Mou}, G., {Yuan}, F., {Bu}, D., {Sun}, M., \& {Su}, M. 2014, \apj, 790, 109

\bibitem[{Paumard} {et~al.}(2006)]{paumard06}
{Paumard}, T., {Genzel}, R., {Martins}, F., {et~al.} 2006, \apj, 643, 1011

\bibitem[{Ponti} {et~al.}(2015)]{Ponti2015}
{Ponti}, G., {Morris}, M.~R., {Terrier}, R., {et~al.} 2015, \mnras, 453, 172

\bibitem[{Ponti} {et~al.}(2019)]{Ponti2019}
{Ponti}, G., {Hofmann}, F., {Churazov}, E., {et~al.} 2019, \nat, 567, 347

\bibitem[{Quataert}(2004)]{Quataert2004}
{Quataert}, E. 2004, \apj, 613, 322

\bibitem[{Rees}(1988)]{Rees1988}
{Rees}, M.~J. 1988, \nat, 333, 523

\bibitem[{Smith} {et~al.}(2001)]{smith2001}
{Smith}, R.~K., {Brickhouse}, N.~S., {Liedahl}, D.~A., \& {Raymond}, J.~C.
  2001, \apjl, 556, L91

\bibitem[{Stolte} {et~al.}(2008)]{stolte08}
{Stolte}, A., {Ghez}, A.~M., {Morris}, M., {et~al.} 2008, \apj, 675, 1278

\bibitem[{Su} {et~al.}(2010)]{su2010}
{Su}, M., {Slatyer}, T.~R., \& {Finkbeiner}, D.~P. 2010, \apj, 724, 1044

\bibitem[{Winter} \& {Lunardini}(2021)]{Winter2021}
{Winter}, W., \& {Lunardini}, C. 2021, Nature Astronomy, 5, 472

\bibitem[{Yalinewich} {et~al.}(2017)]{Yalinewich2017}
{Yalinewich}, A., {Piran}, T., \& {Sari}, R. 2017, \apj, 838, 12

\bibitem[{Yang} {et~al.}(2022)]{Yang2022}
{Yang}, H. Y.~K., {Ruszkowski}, M., \& {Zweibel}, E.~G. 2022, Nature Astronomy,
  6, 584

\bibitem[{Zhang} \& {Guo}(2020)]{ruiyu2020}
{Zhang}, R., \& {Guo}, F. 2020, \apj, 894, 117

\bibitem[{Zhao} {et~al.}(2016)]{Zhao2016}
{Zhao}, J.-H., {Morris}, M.~R., \& {Goss}, W.~M. 2016, \apj, 817, 171

\bibitem[{Zubovas}(2019)]{Zubovas2019}
{Zubovas}, K. 2019, \mnras, 483, 1957

\end{thebibliography}

\end{document}